\documentclass[a4paper]{spie}  
\usepackage[]{graphicx}

\title{Liverpool Telescope 2: a new robotic facility for time domain astronomy in 2020+} 

\author{C.M.~Copperwheat$^a$, I.A.~Steele$^a$, S.D.~Bates$^a$, R.J.~Smith$^a$, M.F.~Bode$^a$, I.~Baker$^b$, T.~Peacocke$^c$ and K.~Thomson$^b$
\skiplinehalf
$^a$Astrophysics Research Institute, Liverpool John Moores University, IC2, Liverpool Science Park, 146 Brownlow Hill, Liverpool, UK \\
$^b$Glyndwr Innovations Ltd., The OpTIC Centre, Ffordd William Morgan, St. Asaph, Wales, UK\\
$^c$Lyncaeus Ltd., 1 George Street, Barnard Castle, Co. Durham, UK\\
}

\authorinfo{Further author information: (Send correspondence to CMC)\\E-mail: c.m.copperwheat@ljmu.ac.uk, Telephone: +44 0151 231 2932}

\begin{document} 
  \maketitle 

\begin{abstract}
The robotic 2m Liverpool Telescope, based on the Canary island of La Palma,  has a diverse instrument suite and a strong track record in time domain science, with highlights including early time photometry and spectra of supernovae, measurements of the polarization of gamma-ray burst afterglows, and high cadence light curves of transiting extrasolar planets. In the next decade the time domain will become an increasingly prominent part of the astronomical agenda with new facilities such as LSST, SKA, CTA and Gaia, and promised detections of astrophysical gravitational wave and neutrino sources opening new windows on the transient universe. To capitalise on this exciting new era we intend to build Liverpool Telescope 2: a new robotic facility on La Palma dedicated to time domain science. The next generation of survey facilities will discover large numbers of new transient sources, but there will be a pressing need for follow-up observations for scientific exploitation, in particular spectroscopic follow-up. Liverpool Telescope 2 will have a 4-metre aperture, enabling optical/infrared spectroscopy of faint objects. Robotic telescopes are capable of rapid reaction to unpredictable phenomena, and for fast-fading transients like gamma-ray burst afterglows. This rapid reaction enables observations which would be impossible on less agile telescopes of much larger aperture. We intend Liverpool Telescope 2 to have a world-leading response time, with the aim that we will be taking data with a few tens of seconds of receipt of a trigger from a ground- or space-based transient detection facility. We outline here our scientific goals and present the results of our preliminary optical design studies.
\end{abstract}


\keywords{Liverpool Telescope, Liverpool Telescope 2, Time domain astronomy, Robotic telescopes, Optical/infrared spectroscopy, Supernovae, Gamma-ray bursts, Gravitational waves}

\section{INTRODUCTION}
\label{sec:intro}  

The Liverpool Telescope\cite{Steele04} (LT) is a 2 metre aperture telescope located at the Observatorio del Roque de Los Muchachos on the Canary island of La Palma, owned and operated by Liverpool John Moores University. The Liverpool Telescope is fully robotic and completely autonomous throughout the night, and is the largest robotic telescope dedicated to scientific observing in operation. Robotic telescopes are ideal tools for the study of the time domain, since they allow flexible monitoring of variable sources on a wide range of observing cadences, and are capable of the most rapid reactions to unpredictable transient phenomena. One of the most active areas of astronomical research today is the study of transient phenomena such as supernovae (SNe) and gamma-ray bursts (GRBs). The operational life of the LT has coincided with the beginnings of the era of wide field `synoptic' surveys such as the Palomar Transient Factory (PTF)\cite{Rau09} as well as the Swift (GRB) Mission\cite{Gehrels04}, and the rapid reaction of the LT has enabled it to take a leading role in transient follow-up. The LT benefits from a diverse instrument suite, all of which are simultaneously mounted, allowing a flexible observing strategy for individual sources. One illustrative science highlight has been the measurements of the magnetised jets of emission in early-time GRB afterglows detected with the novel RINGO series of imaging polarimeters\cite{Mundell07,Steele09,Mundell13}. These observations provide an example of scientific results which would be unachievable without the rapid reaction and instrumental capability of the LT.

Interest in the time domain is set to escalate in the coming decades with the advent of a number of major new facilities. Optical transient astronomy will be revolutionised by facilities such as the Large Synoptic Survey Telescope\cite{Ivezic08}. (LSST), which will image the entire Southern sky every few nights and is expected to issue $\sim$$1 \times 10^6$ alerts per night, of which around $10,000$--$100,000$ will be new explosive transients. Other new facilities such as the Square Kilometre Array\cite{Carilli04} (SKA) and the Cherenkov Telescope Array\cite{Actis11} (CTA) will open the temporal window on new regions of the electromagnetic spectrum. We are also entering into an era when astrophysical objects might be routinely detected by non-electromagnetic means. The IceCube\cite{Karle03} detector in Antarctica is engaged in searches for cosmic neutrinos, and this will be joined by KM3Net in the Northern hemisphere\cite{Ulrich14}. In 2015 the Advanced LIGO \cite{Abbott09,Harry10} and Advanced Virgo\cite{Degallaix13} detectors will begin science operations, searching for gravitational wave emission from coalescing neutron star and black hole binaries. The challenge for astronomers will be to detect the electromagnetic counterparts to these violent events.

The new era of time domain astronomy calls for a new generation of follow-up telescope for scientific exploitation. The next generation of optical survey facilities such as LSST will not only report transients at much fainter magnitudes, but also the cadences of these large surveys means that much of the long term photometric monitoring currently provided by follow-up telescopes will be provided `for free' by the survey itself. In Autumn 2012 Liverpool John Moores University therefore began a feasibility study for a successor to the LT, with the aim that Liverpool Telescope 2 (LT2) will come into operation around the beginning of the next decade, which coincides nicely with the commencement of operations of many major new time domain facilities. In this paper we report our progress to date. In Section \ref{sec:science} we provide a brief overview of the main scientific drivers of the project. This is a summary of a more detailed review which we will publish this year\cite{Copperwheat14}. In Section \ref{sec:design} we discuss the key technical drivers. Our preliminary design studies have concentrated on the telescope optics, and we give an overview of our results in Section \ref{sec:optics}. Finally in Section \ref{sec:site} we briefly discuss our site choice and options for the future of the existing LT.

\section{MAIN SCIENCE DRIVERS}
\label{sec:science}  

In this section we detail the main science areas which drive the case for LT2. For reasons of space we limit ourselves to three topics, but we note that there are many other new time domain facilities coming into operation over the next decade for which robotic optical/infrared follow-up or simultaneous observing could be extremely productive. As well as facilities across the electromagnetic spectrum such as CTA and SKA, the next generation of transiting exoplanet-finders will focus on bright stars to maximise the scope for ground-based follow-up\cite{Ricker10,Wheatley13,Rauer13}. The final Gaia\cite{Perryman01} catalogue will also be published in $2020$, and will contain an unprecedented number of variable object detections.

\subsection{Supernovae}

\begin{figure}
\centering
\includegraphics[angle=270,width=0.5\columnwidth]{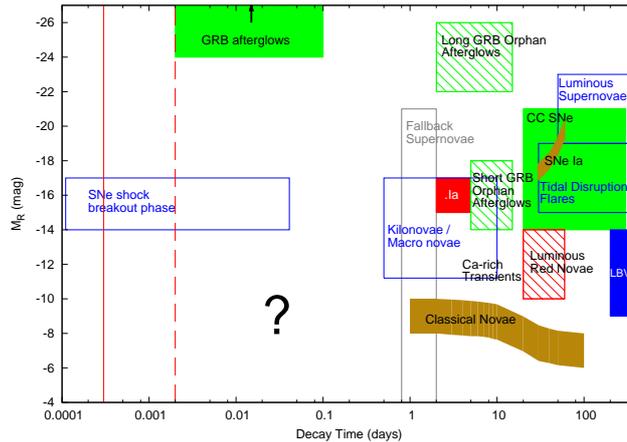}
\hfill
\caption{Explosive transient phase space. This is a modified version of a figure which appears in the LSST Science Book\cite{Lsst09,Rau09}, which shows the unexplored sub-day phase space which will be probed by the next generation of synoptic surveys. We include here the SNe shock breakout phase as the example of a phenomenon which will continue to be of great interest and will require a sub-hour follow-up response. The vertical red lines denote for reference the typical response time of the LT (dashed) and LT2 (solid).} \label{fig:transients} \end{figure}

SNe are of intrinsic interest as spectacular and catastrophic events which mark the end of the life of the star, and SNe Ia in particular are of great utility as standardizable candles for the measurement of cosmic distances. However there are still a great many unknowns. The discovery of `super-Chandrasekhar' SNe Ia with luminosities of about twice the average\cite{Astier06,Howell06} challenges the traditional single-degenerate formation scenario. It is quite possible at this stage that there are alternative routes to a SN Ia explosion, which could have significant consequences for the use of SNe Ia as standard candles. By comparison core collapse SNe consist of a wide range of subclasses based on observed differences in the light curves or spectra of the aftermath of the explosion. The difficulty is linking these observed differences with differences in the physical parameters of the progenitors, and effects such as the initial mass, metallicity, rotation rate and the presence and properties of binary companions or magnetic fields. The main aim for both SNe Ia and core collapse SNe is establishing a more statistically complete sample. This challenge is met by current wide field synoptic surveys such as PTF, Skymapper and Pan-STARRS, with LSST expected to revolutionise the field in the next decade. Wide field surveys are uncovering a wealth of transient phenomena that would mostly be missed by the previous search mode, which aimed mostly at bright galaxies and had cadences optimised mostly for SNe Ia. This is illustrated in Figure \ref{fig:transients}, which is a version of a figure originally created to show the explosive transient phase space\cite{Rau09}, later modified to highlight the unexplored `fast and faint' regime which will be probed by LSST\cite{Lsst09}.

There is however a pressing need for spectroscopic classification and follow-up of the wide field surveys. Historically only $\sim$$10$ per cent of PTF transients receive a classification, and while this is set to improve with the advent of new low resolution spectrographs (such as the `SED machine' on the Palomar 60-inch telescope\cite{Ngeow13}, and SPRAT on the LT), the need will become much more acute in the LSST era. Recently the PESSTO survey has demonstrated that dedicating large amounts of 4-metre time to spectroscopic follow-up can be extremely productive, and the flexible scheduling capability of robotic telescopes makes it straightforward to obtain well-sampled observations of large numbers of targets, and a rapid reaction capability can be very important for SNe discovered very soon after the explosion. For example the first spectrum of SN2011fe\cite{Nugent11} was obtained with the LT, and the detection of unburned carbon and high velocity oxygen was the first decisive evidence that CO white dwarfs are SN Ia progenitors. 

\subsection{Gamma-ray Bursts}

Gamma-ray bursts (GRBs) are the most energetic explosions to be detected in the
universe. They are characterised by a short and intense burst of prompt gamma-ray emission which is commonly followed by a rapidly fading afterglow detectable at X-ray and optical wavelengths. The Swift Gamma-Ray Burst Mission and the Fermi Gamma Ray Space Telescope continuously scan a large fraction of the sky for gamma-ray events and provide real-time triggers to ground-based telescopes for follow-up of the afterglows. GRBs can be divided into long/soft and short/hard events, depending on the duration of the prompt emission (with the cut at $\sim$$2$ sec) and spectral hardness. The long events are thought to be associated with the death of massive stars\cite{Woosley93,MacFadyen99,Iwamoto98,Mazzali03,Mazzali05b} while the latter may be due to the merging of two compact objects, such as neutron stars or stellar mass black holes\cite{Eichler99}. However we continue to find unusual phenomena within the population, for example ultra-long GRBs (ulGRBs), with prompt phases lasting for $\sim$$10,000$ seconds\cite{Gendre13}. As some of the most distant objects known, high redshift GRBs have great potential as probes of cosmological parameters, reionisation and the star formation history of the universe, and as a means of directly detecting the first stars in the universe. However there are many open questions regarding the physical processes at work during the initial, prompt phase of the GRB, in terms of particle acceleration and radiation processes, as well as the nature of the compact merging components which cause the short events..

The rapid reaction capability of the LT, coupled with a diverse instrument suite and a flexible real-time pipeline which makes automated decisions as to the type of data to obtain (multicolour photometry,
spectroscopy and/or polarimetry) based on the brightness of the afterglow\cite{Guidorzi06}, has made it an extremely productive facility for GRB follow-up. The typical reaction time of the LT to a trigger from a spaced-based facility is $2$--$3$ min. The reaction time is the crucial parameter for GRB follow-up. Since the decay rate of the afterglows is extremely rapid, in many cases a telescope with a greater slew speed can collect more photons than a telescope with a significantly larger aperture. The case for GRB science in the next decade is however predicated on the assumption that a space-based triggering system will be in existence at that time. Both Swift (launched 2004) and Fermi (launched 2008) are already in the extended phases of their missions. Numerous successor missions have however been discussed, the most promising of which is the joint French-Chinese mission SVOM\cite{Gotz09}.

\subsection{Gravitational wave counterparts}

In 2015 the advanced versions of the LIGO\cite{Abbott09,Harry10} and Virgo\cite{Degallaix13} detectors will begin operations, building up to full sensitivity in $\sim$$2022$. The best sensitivities will be for GW signal frequencies of $\sim$$100$ -- $200$ Hz\cite{Aasi13}, with the most likely detections in this frequency range being coalescing binary systems with neutron star or black hole components. It is difficult to estimate the detection rate of such mergers, but a `realistic rate' of $40$/year is predicted, although the actual rate could be one order of magnitude higher or two lower \cite{Abadie10}. The importance of a significant gravitational wave detection is enormous, and would open an entirely new window onto the universe. The detection of an electromagnetic counterpart to any event will be important for verification as well as scientific exploitation. However, the nature of the counterpart is currently not well understood. It is unclear what fraction of mergers will be accompanied by a short-GRB, and even for those that are, the prompt emission will only be detected for the small fraction in which the opening angle of the jet encompasses the line of sight. The detected emission is therefore likely to be entirely due to the interaction of the relativistic outflow with the surrounding medium\cite{VanEerten11,Metzger12}. Evidence has been detected for `kilonova' emission in the aftermath of GRB 130603B\cite{Berger13,Tanvir13}, which is a red transient powered by the radioactive decay of heavy nuclei synthesised in the neutron-rich merger ejecta. This is a promising candidate for the gravitational wave counterpart.

A significant additional challenge is the large positional uncertainty of any GW detection, which will be of the order of a few square degrees at best\cite{Aasi13b}. A field of this size will contain large numbers of contaminating transients, such as near Earth objects and variable stars. It is likely that initial detection will rely on arrays of small, dedicated telecopes; or wide field survey facilities such as LSST. The role of conventional telescopes is more likely to be focused on scientific exploitation of the counterpart once it is discovered by wide field facilities, and so given that the counterpart may potentially be fast-fading, a rapid reaction capability is again important. The LT was part of the follow-up programme for the previous generation of detectors and demonstrated the benefits of a large aperture robotic telescope for work of this nature: the images obtained with the RATCam instrument $0.35$ days after candidate event G23004 provided the first real constraints on any kilonova emission\cite{Abadie12b, Aasi13}.

\section{TELESCOPE DESIGN}
\label{sec:design}

\subsection{Aperture}

Currently a typical task for the LT (e.g. an interesting supernova detection) involves photometric monitoring of the light from the source on timescales of hours, days and weeks: the rise, peak and decay which characterises the explosion. As we have noted, in the era of LSST much of this monitoring will be provided ‘for free’ by the survey telescope, since it will be imaging the entire Southern sky on timescales of a few days. What will be missing is follow-up spectroscopy to characterise the abundances and velocities of the components in the ejecta. The $2$-metre aperture of the LT is insufficient for spectroscopic follow-up in the era of LSST, which will report transients down to $24$th magnitude. We therefore intend LT2 to have an aperture of 4 metres diameter. For this aperture an $1800$ second integration through an $R=1000$ spectrograph would provide a S/N of greater than $10$ for a $20$th magnitude object in the optical $R$-band, or an $18$th magnitude object in the infrared $J$-band. This is at the bright end of the LSST transient space, but the expected number of transient alerts from LSST is so large that there will still be vast amounts of targets for us to choose from.

\subsection{Rapid response}

The next decade will see a vast increase in the capabilities of survey facilities, in terms of sky coverage, cadence and wavelength, revealing new transient and variable objects at an unprecedented rate. The opening of the time domain window on new electromagnetic and non-electromagnetic regimes, and also the exploration of the faint/fast region of the transient magnitude / timescale phase space (Figure \ref{fig:transients}) will provide new insights into existing phenomena, but also discover new classes of rapidly varying objects. A recent interesting phenomenon for example are the millisecond radio bursts which may be associated with neutron star - neutron star mergers\cite{Lorimer07,Thornton13}. Clearly as we push down to faster events the reactive capabilities of our follow-up telescopes are crucial, and so we aim to provide a telescope with a world-leading response time in order to cover an area of the parameter space which is unfilled by existing follow-up facilities. Like LT, LT2 will therefore be a fully robotic telescope, as the LT has demonstrated the power of robotic operations for the study of targets of opportunity. We have made some estimations of what we believe to be possible in terms of telescope slew time and have set the following target: we aim that LT2 will, on average, be able to start obtaining data within $30$ seconds of receipt of an electronically transmitted alert from another facility. This time incorporates the telescope blind pointing time, the mirror settling time and any mechanical movement of the enclosure, although with respect to this last point the enclosure will likely be of the same clamshell design as the LT, which does not require any movement through the night. We constructed a simple model to investigate the required angular velocities and accelerations of the telescope axes to meet this requirement, and find the requirements are relatively modest ($v \sim 4$deg/s and $a \sim 0.4$deg/s/s), although the ease  with which this can be achieved is dependent on the moment of inertia of the telescope. Potentially the most challenging part of the mechanical design is the mirror support, which will need to maintain the mirror alignment under these accelerations as well as enable a very short settling time.

\subsection{Instrumentation}

The diverse instrument suite of the LT is one of its core strengths: all instruments are mounted simultaneously, and instrument changes can be made in the middle of a night with a low ($< 1$ minute) overhead. LT instrumentation can therefore also be fairly specialised and relatively simple, which means the lead time from instrument concept to science operations is quite low. This has enabled the LT to respond quickly to new and evolving scientific needs. We would like LT2 to have these same strengths, and therefore a requirement of the telescope design is the ability to mount up to five instruments simultaneously, with the capability for automatic changes. We anticipate that the first-light instrument will be an optical/infrared spectrograph capable of low to intermediate dispersions (up to $R$$\sim$few thousand). The wavelength range of the spectrograph is to be decided, although we would aim to push as far into the infrared as possible to facilitate the study of extragalactic transients. Our design specification for the telescope itself imposes an effective wavelength range of at least $350$nm to $2.0\mu$m, covering the optical U-band to the infrared H-band. Since our focus is on spectroscopy we also have fairly modest requirements for field of view and image quality. Our minimum requirements are for a field of view of $15'$ diameter, and an image quality such that at times of median seeing ($\sim$$0.8''$ at La Palma) the image quality on-axis from the full optical system must be better than $1.0''$ diameter measured at $80$ per cent encircled energy, and better than $4.0''$
diameter measured at $99\%$ encircled energy. These minimum requirements may well be improved upon in the final design.

\section{OPTICS}
\label{sec:optics}

\begin{figure}
\centering
\includegraphics[angle=0,width=0.5\columnwidth]{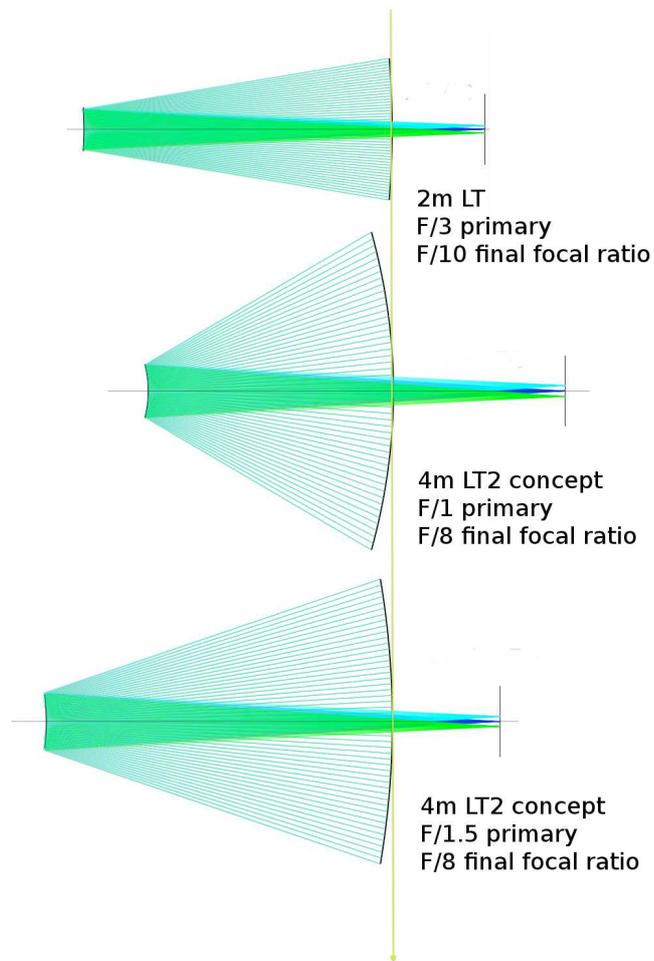}
\hfill
\caption{Comparison of potential LT2 optical layouts to the layout of the LT, at the same scale.} \label{fig:mirrors} \end{figure}

The design studies we have undertaken to date have concentrated on the optics, since the mass and the layout of the optical components will drive the mechanical design. These studies recommend a Ritchey-Chr\'etien with a f/1.0 -- f/1.5 primary, and we have considered a range of final focal ratios between f/6.5 and f/10. For comparison the LT is an f/10 Ritchey-Chr\'etien with an f/3.0 primary mirror. In Figure \ref{fig:mirrors} we compare the two potential f/8 optical layouts to the layout of the LT. The separation between the two mirrors is comparable due to the faster primaries used for the LT2 designs. The different final focal lengths represent different approaches to the problem of minimising the moment of intertia in order to maximise potential slew speed. A short final focal length reduces the overall length of the system, whereas a long final focal length allows a reduction of the size and hence mass of the secondary, which can more than compensate for the increased length of the system. In general, the designs with an f/1.0 primary seem preferable, since they have a larger (3.0 metre vs 1.5 metre) back focal distance making them more suitable for Nasmyth than Cassegrain use. The f/1.0 primary with an f/8 final focal length provides an acceptable plate scale. All the designs meet our image quality and field of view specification on  a curved focal plane, but we find a simple single element field flattener will allow good image quality to the edge of the $15'$ field. 

\begin{table}
\caption{Segment parameters for a 6, 18 and 36 segment primary mirror.}
\label{tab:params}
\begin{center}
\begin{tabular}{cccccccc}
Segment & Dimension 		& Dimension 	& Edge 		& Centre 	& Mass 	& No. 		& Print\\
no. 	& across corners 	& across flats 	& thickness 	& thickness 	& (kg) 	& support 	& through error\\
 	& (mm) 			& (mm) 		& (mm) 		& (mm) 		&  	& points 	& (nm)\\
\hline
6	& 1819			& 1582		& 113.4		& 65.2		& 454.4	& 27		& 59.65\\
18	& 1041			& 903		& 53.9		& 37.7		& 77.5	& 18		& 29.28\\
36	& 734			& 636		& 34.5		& 26.6		& 25.8	& 18		& 22.71\\	
\hline
\end{tabular}
\end{center}
\end{table}

\begin{figure}
\centering
\includegraphics[angle=0,width=0.5\columnwidth]{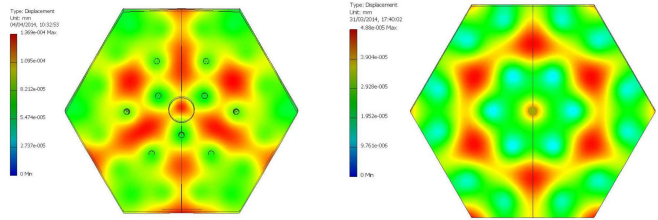}
\hfill
\caption{Print through for the $6$ segment design (left) and $18$ segment design (right) the peak-to-valley values are $59.65$ and $29.28$nm, respectively.} \label{fig:printthrough} \end{figure}

The requirement for rapid slewing necessitates a light-weight design for the optics. We estimate a thin Zerodur meniscus F/10 primary would have a mass of approximately $5500$kg. We have therefore been exploring options for a segmented telescope mirror, which substantially reduces the mass. We summarise our results in Table \ref{tab:params}, in which we give the dimensions and mass of hexagonal segments which would be used to make a $6$, $18$ or $36$ piece primary mirror with a total collecting area of $4\pi$m$^2$. We assume the mirror support would consist of the lateral support system housed within the volume of the optic, along the lines of the method which was developed for the Keck telescopes and which will be used for the ESO E-ELT. The support system drives the thickness and maximum sagitta of the optic. For the three segment designs we list in the table the number of support points per segment and the peak-to-valley print through error, which we also plot for the $6$ and $18$ segment design in Figure \ref{fig:printthrough}. The worst print through error is for the $6$ segment design, but this is still small and comparable in size to the value calculated for the E-ELT segments.
 
\begin{figure}
\centering
\includegraphics[angle=0,width=0.5\columnwidth]{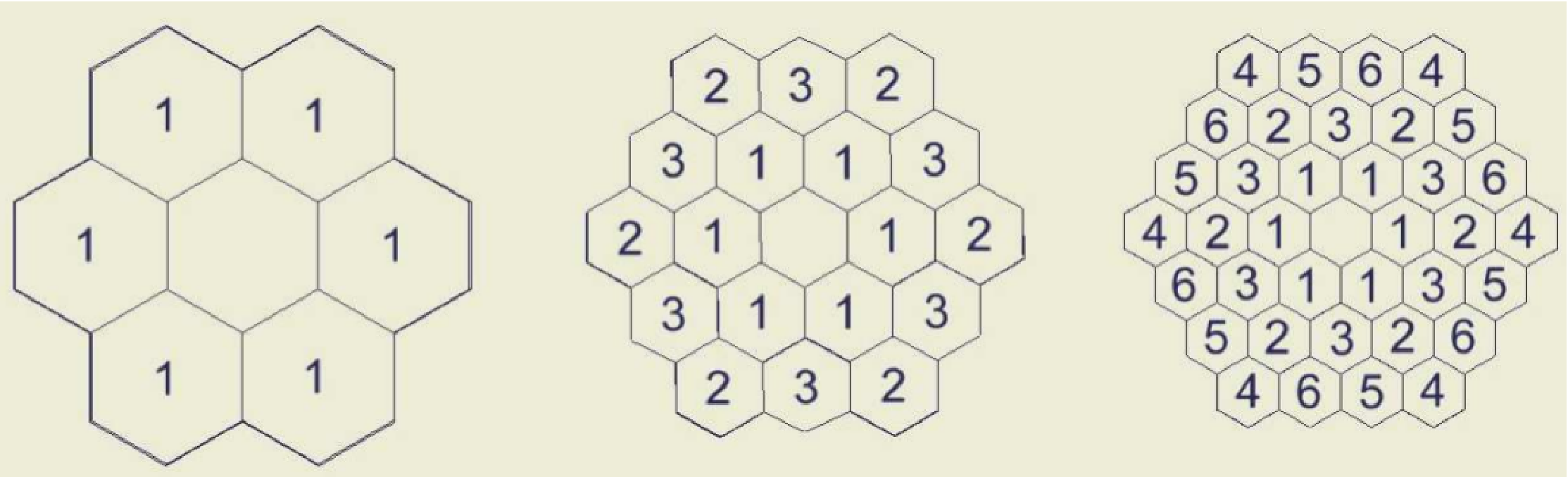}
\hfill
\caption{Family groups for the different segmented options.} \label{fig:family} \end{figure}

The total mirror masses are $\sim$$2700$, $1400$ and $920$kg for the $6$, $18$ and $36$ segment mirror respectively. The $18$ and $36$ segment designs are more attractive from an ease-of-handling point of view, which impacts on operations costs. However, our initial cost analysis suggests the cost of manufacture is very sensitive to the polishing and testing cost, which rises substantially as the number of family groups is increased (Figure \ref{fig:family}), which is a particular problem for the $36$ segment option. On balance then we tend to favour the $18$ segment design at this stage.

A technical complication for a segmented mirror is alignment of the segments. At the very least the mirror support system must co-align the segments such that their individual images overlap and are in mutual focus. However, the ideal is an alignment which maintains optical phase coherence across the mirror surface. There is a factor  $\sim$$100$ difference in the required alignment tolerances between these two extremes, and so maintaining the mirror in a phased alignment represents a significant technical challenge. If operating un-phased, the effective resolution of the telescope is defined by the diffraction limit of the individual segments. However, since we envisage this telescope will be primarily employed for spectroscopic work, maximising image quality is not the highest priority. For the $18$ segment design for example, the angular resolution of a single segment would be $0.17''$ at the optical R-band ($650$nm). Given that we do not plan to deploy an adaptive optics system, the point spread function (PSF) will therefore be dominated by atmospheric effects at these wavelengths. However the un-phased design does limit image quality at infrared wavelengths: the angular resolution of the $18$ segment design would be $0.53''$ at $2\mu$m. This improves to $0.14''$ if we use the larger segments of the $6$ segment design. 

An additional complication to be considered in an un-phased design is the nature of the point spread function (PSF) from a hexagonal mirror, which will contain six strong diffraction spikes. We are currently performing a numerical analysis of the total energy scattered into the wings investigating the effect of an un-phased mirror on spectroscopy, where you are effectively working with many monochromatic images of the slit and hence interference effects are likely to be greater. Alternative (non-hexagonal) tiling schemes could be investigated, for example a larger hexagon can be created from 6 pentagons and a central smaller hexagon, making a 42 segment primary.  The advantage of such a scheme might be in reducing any reinforcement of diffraction spikes caused by the traditional purely hexagonal design.

\section{SITE, AND FUTURE OF LT}
\label{sec:site}

In partnership with the Instituto de Astrofísica de Canarias we intend to construct LT2 at the Observatorio del Roque de Los Muchachos (ORM) on the Canary island of La Palma, Spain: the same site as the LT. The ORM is one of the best observing sites in the world, with low levels of light pollution, median seeing of $0.8''$, and a high proportion of photometric nights. Our familiarity with the site brings obvious logistical advantages, but the latitude of the Canary Islands makes them an ideal site for follow-up of transients discovered with both Northern and Southern survey facilities.

Co-locating LT2 with LT opens up a couple of interesting possibilities for the future of the LT itself. One option we are considering is to convert the LT into a dedicated wide field imaging telescope with the addition of a new prime focus camera. The two telescopes would then be run as a combined facility. The facility could provide simultaneous spectroscopy and photometry of the same target and the LT would be operated as our own transient finding facility. PTF has demonstrated the utility of this factory approach to transient science, and a fully automated robotic facility could be extremely productive. 

\begin{figure}
\centering
\includegraphics[angle=0,width=0.5\columnwidth]{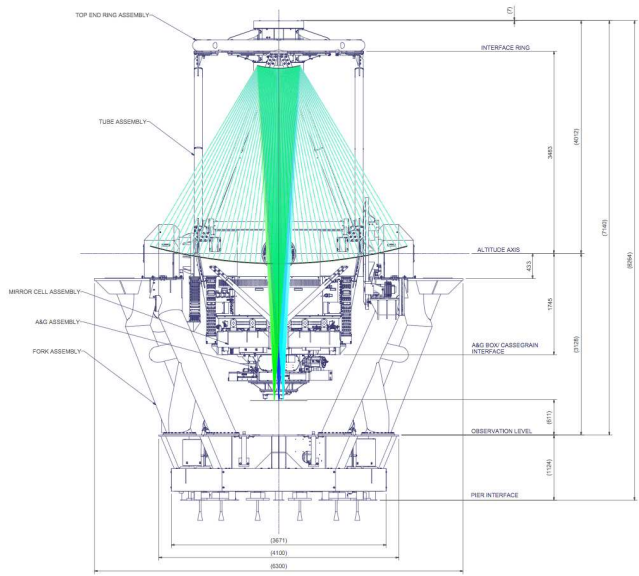}
\hfill
\caption{LT2 optical path (f/1 primary, f/8 final focal ratio) superimposed on the LT at same scale.} \label{fig:upgrade} \end{figure}

Alternatively, the compact nature of our optical designs opens the possibility for LT2 to fit within the current LT enclosure. Figure \ref{fig:upgrade} shows that the space envelope required for the optics alone (making no allowance for the telescope structure) fits within the current LT space envelope, although the exact positioning of the optical elements will require a detailed design study. In the layout presented, the primary mirror has been placed as high as possible in order to reduce vignetting from the edge of the telescope enclosure thus optimising access to the LSST field, which as presented should be similar to the LT. For the $36$ segment option, the total mass of the optical elements in LT2 would be less than the 2-metre LT mirror, which implies the existing concrete telescope pier could be reused. This arrangement shown does also appear to allow potential reuse of the telescope fork assembly. Re-using parts of the existing infrastructure would significantly simplify the project, and so LT2 could potentially come into operation sooner. The price of course would be the loss of the existing $2$-metre facility, as well as an operational downtime of $\sim$$1$ year. As we progress into the mechanical design stage we will aim to keep this idea a viable option.

\section{SUMMARY}
\label{sec:summary}

The Astrophysics Research Institute at Liverpool John Moores University intends to build a 4-metre class fully robotic telescope to be co-located with the existing Liverpool Telescope on the Canary island of La Palma. We intend operations to begin in the early years of the next decade. This telescope will be dedicated to time domain science, in particular the exploitation of explosive transients, and it will feature a variety of simultaneously mounted instrumentation for flexible follow-up. One first light instrument will be an intermediate resolution, optical/infrared spectrograph. The telescope will be capable of extremely rapid reaction to targets-of-opportunity, with the aim that it will be taking data within $30$ seconds of receipt of a trigger from a survey facility. This will necessitate a lightweight telescope design, and with that in mind our initial design studies have concentrated on options for the primary mirror. A segmented primary consisting of $6$ or $18$ segments seems a viable option, resulting in a total mirror mass which is similar to the monolithic $2$-metre mirror currently in use on the LT. We are currently conducting further studies on the issue of segment alignment, and are about to begin to address the mechanical design. Ideally we would aim to operate LT2 alongside LT as a combined robotic facility, however we note that the compact nature of our optical layouts make possible an alternative upgrade path to LT2 in which LT is decomissioned and the enclosure, pier and possibly even parts of the fork assembly are recycled.

\acknowledgments 

The Liverpool Telescope is operated on the island of La Palma by Liverpool John Moores University in the Spanish Observatorio del Roque de los Muchachos of the Instituto de Astroﬁsica de Canarias with financial support from the UK Science and Technology Facilities Council.


\bibliography{lt2}   

\begin{thebibliography}{10}

\bibitem{Steele04}
{Steele}, I.~A., {Smith}, R.~J., {Rees}, P.~C., {Baker}, I.~P., {Bates}, S.~D.,
  {Bode}, M.~F., {Bowman}, M.~K., {Carter}, D., {Etherton}, J., {Ford}, M.~J.,
  {Fraser}, S.~N., {Gomboc}, A., {Lett}, R.~D.~J., {Mansfield}, A.~G., and
  {Marchant}, J.~M., ``{The Liverpool Telescope: performance and first
  results},'' in [{\em Society of Photo-Optical Instrumentation Engineers
  (SPIE) Conference Series}{\nolinebreak\hspace{0.1em}]},  {Oschmann}, Jr.,
  J.~M., ed., {\em Society of Photo-Optical Instrumentation Engineers (SPIE)
  Conference Series} {\bf 5489},  679--692 (Oct. 2004).

\bibitem{Rau09}
{Rau}, A., {Kulkarni}, S.~R., {Law}, N.~M., {Bloom}, J.~S., {Ciardi}, D.,
  {Djorgovski}, G.~S., {Fox}, D.~B., {Gal-Yam}, A., {Grillmair}, C.~C.,
  {Kasliwal}, M.~M., and {Nugent}, P.~E., ``{Exploring the Optical Transient
  Sky with the Palomar Transient Factory},'' {\em PASP}~{\bf 121},  1334--1351
  (Dec. 2009).

\bibitem{Gehrels04}
{Gehrels}, N., {Chincarini}, G., {Giommi}, P., {Mason}, K.~O., {Nousek}, J.~A.,
  {Wells}, A.~A., {White}, N.~E., {Barthelmy}, S.~D., {Burrows}, D.~N.,
  {Cominsky}, L.~R., {Hurley}, K.~C., {Marshall}, F.~E., {M{\'e}sz{\'a}ros},
  P., {Roming}, P.~W.~A., {Angelini}, L., {Barbier}, L.~M., and {Belloni}, T.,
  ``{The Swift Gamma-Ray Burst Mission},'' {\em ApJ}~{\bf 611},  1005--1020
  (Aug. 2004).

\bibitem{Mundell07}
{Mundell}, C.~G., {Steele}, I.~A., {Smith}, R.~J., {Kobayashi}, S., {Melandri},
  A., {Guidorzi}, C., {Gomboc}, A., {Mottram}, C.~J., {Clarke}, D.,
  {Monfardini}, A., {Carter}, D., and {Bersier}, D., ``{Early Optical
  Polarization of a Gamma-Ray Burst Afterglow},'' {\em Science}~{\bf 315},
  1822-- (Mar. 2007).

\bibitem{Steele09}
{Steele}, I.~A., {Mundell}, C.~G., {Smith}, R.~J., {Kobayashi}, S., and
  {Guidorzi}, C., ``{Ten per cent polarized optical emission from GRB090102},''
  {\em Nature}~{\bf 462},  767--769 (Dec. 2009).

\bibitem{Mundell13}
{Mundell}, C.~G., {Kopa{\v c}}, D., {Arnold}, D.~M., {Steele}, I.~A., {Gomboc},
  A., {Kobayashi}, S., {Harrison}, R.~M., {Smith}, R.~J., {Guidorzi}, C.,
  {Virgili}, F.~J., {Melandri}, A., and {Japelj}, J., ``{Highly polarized light
  from stable ordered magnetic fields in GRB120308A},'' {\em Nature}~{\bf 504},
   119--121 (Dec. 2013).

\bibitem{Ivezic08}
{Ivezic}, Z., {Tyson}, J.~A., {Acosta}, E., {Allsman}, R., {Anderson}, S.~F.,
  {Andrew}, J., {Angel}, R., {Axelrod}, T., {Barr}, J.~D., {Becker}, A.~C.,
  {Becla}, J., {Beldica}, C., {Blandford}, R.~D., {Bloom}, J.~S., {Borne}, K.,
  and {Brandt}, W.~N., ``{LSST: from Science Drivers to Reference Design and
  Anticipated Data Products},'' {\em ArXiv e-prints}  (May 2008).

\bibitem{Carilli04}
{Carilli}, C.~L. and {Rawlings}, S., ``{Science with the Square Kilometre
  Array},'' {\em NewAR}~{\bf 48},  979--1606 (Dec. 2004).

\bibitem{Actis11}
{Actis}, M., {Agnetta}, G., {Aharonian}, F., {Akhperjanian}, A., {Aleksi{\'c}},
  J., {Aliu}, E., {Allan}, D., {Allekotte}, I., {Antico}, F., {Antonelli},
  L.~A., and et~al., ``{Design concepts for the Cherenkov Telescope Array CTA:
  an advanced facility for ground-based high-energy gamma-ray astronomy},''
  {\em Experimental Astronomy}~{\bf 32},  193--316 (Dec. 2011).

\bibitem{Karle03}
{Karle}, A., {Ahrens}, J., {Bahcall}, J.~N., {Bai}, X., {Becka}, T., {Becker},
  K.-H., {Besson}, D.~Z., {Berley}, D., {Bernardini}, E., {Bertrand}, D.,
  {Binon}, F., {Biron}, A., {B{\"o}ser}, S., {Bohm}, C., {Botner}, O., and
  {Bouhali}, O., ``{IceCube - the next generation neutrino telescope at the
  South Pole},'' {\em Nuclear Physics B Proceedings Supplements}~{\bf 118},
  388--395 (Apr. 2003).

\bibitem{Ulrich14}
{Ulrich F.~Katz for the KM3NeT Collaboration}, ``{News from KM3NeT},'' {\em
  ArXiv e-prints}  (Mar. 2014).

\bibitem{Abbott09}
{Abbott}, B.~P., {Abbott}, R., {Adhikari}, R., {Ajith}, P., {Allen}, B.,
  {Allen}, G., {Amin}, R.~S., {Anderson}, S.~B., {Anderson}, W.~G., {Arain},
  M.~A., and et~al., ``{LIGO: the Laser Interferometer Gravitational-Wave
  Observatory},'' {\em Reports on Progress in Physics}~{\bf 72},  076901 (July
  2009).

\bibitem{Harry10}
{Harry}, G.~M. and {LIGO Scientific Collaboration}, ``{Advanced LIGO: the next
  generation of gravitational wave detectors},'' {\em Classical and Quantum
  Gravity}~{\bf 27},  084006 (Apr. 2010).

\bibitem{Degallaix13}
{Degallaix}, J., {Accadia}, T., {Acernese}, F., {Agathos}, M., {Allocca}, A.,
  {Astone}, P., {Ballardin}, G., {Barone}, F., {Bejger}, M., {Beker}, M.~G.,
  {Bitossi}, M., and {Bizouard}, M.~A., ``{Advanced Virgo Status},'' in [{\em
  9th LISA Symposium}{\nolinebreak\hspace{0.1em}]},  {Auger}, G.,
  {Bin{\'e}truy}, P., and {Plagnol}, E., eds., {\em Astronomical Society of the
  Pacific Conference Series} {\bf 467},  151 (Jan. 2013).

\bibitem{Copperwheat14}
{Copperwheat}, C.~M., {Steele}, I.~A., M., B.~R., {Bates}, S.~D., {Bersier},
  D., {Bode}, M.~F., {Carter}, D., {Collins}, C.~A., {Darnley}, M., {Davis},
  C.~J., {Harman}, D.~J., {James}, P.~A., {Kobayashi}, S., {Marchant}, J.~M.,
  {Mazzali}, P.~A., {Mottram}, C.~J., {Mundell}, C.~G., {Newsam}, A., and
  {Smith}, R.~J., ``{Liverpool Telescope 2: a new robotic facility for rapid
  transient follow-up},'' {\em In prep.}  (2014).

\bibitem{Ricker10}
{Ricker}, G.~R., {Latham}, D.~W., {Vanderspek}, R.~K., {Ennico}, K.~A.,
  {Bakos}, G., {Brown}, T.~M., {Burgasser}, A.~J., {Charbonneau}, D.,
  {Clampin}, M., {Deming}, L.~D., {Doty}, J.~P., {Dunham}, E.~W., {Elliot},
  J.~L., and {Holman}, M.~J., ``{Transiting Exoplanet Survey Satellite
  (TESS)},'' in [{\em American Astronomical Society Meeting Abstracts
  215}{\nolinebreak\hspace{0.1em}]},  {\em Bulletin of the American
  Astronomical Society} {\bf 42},  450 (Jan. 2010).

\bibitem{Wheatley13}
{Wheatley}, P.~J., {Pollacco}, D.~L., {Queloz}, D., {Rauer}, H., {Watson},
  C.~A., {West}, R.~G., {Chazelas}, B., {Louden}, T.~M., {Walker}, S.,
  {Bannister}, N., {Bento}, J., {Burleigh}, M., {Cabrera}, J., and
  {Eigm{\"u}ller}, P., ``{The Next Generation Transit Survey (NGTS)},'' in
  [{\em European Physical Journal Web of
  Conferences}{\nolinebreak\hspace{0.1em}]},  {\em European Physical Journal
  Web of Conferences} {\bf 47},  13002 (Apr. 2013).

\bibitem{Rauer13}
{Rauer}, H., ``{The PLATO 2.0 mission},'' {\em European Planetary Science
  Congress 2013, held 8-13 September in London, UK, id.EPSC2013-707}~{\bf 8},
  707 (Sept. 2013).

\bibitem{Perryman01}
{Perryman}, M.~A.~C., {de Boer}, K.~S., {Gilmore}, G., {H{\o}g}, E.,
  {Lattanzi}, M.~G., {Lindegren}, L., {Luri}, X., {Mignard}, F., {Pace}, O.,
  and {de Zeeuw}, P.~T., ``{GAIA: Composition, formation and evolution of the
  Galaxy},'' {\em \aap}~{\bf 369},  339--363 (Apr. 2001).

\bibitem{Lsst09}
{Abell}, P.~A., {Allison}, J., {Anderson}, S.~F., {Andrew}, J.~R., {Angel},
  J.~R.~P., {Armus}, L., {Arnett}, D., {Asztalos}, S.~J., and {Axelrod}, T.~S.
  e.~a., ``{LSST Science Book, Version 2.0},'' {\em ArXiv e-prints}~{\bf
  {arXiv:0912.0201}} (Dec. 2009).

\bibitem{Astier06}
{Astier}, P., {Guy}, J., {Regnault}, N., {Pain}, R., {Aubourg}, E., {Balam},
  D., {Basa}, S., {Carlberg}, R.~G., {Fabbro}, S., {Fouchez}, D., {Hook},
  I.~M., {Howell}, D.~A., {Lafoux}, H., {Neill}, J.~D., and
  {Palanque-Delabrouille}, ``{The Supernova Legacy Survey: measurement of
  {$\Omega$}$_{M}$, {$\Omega$}$_{Λ}$ and w from the first year data set},''
  {\em A\&A}~{\bf 447},  31--48 (Feb. 2006).

\bibitem{Howell06}
{Howell}, D.~A., {Sullivan}, M., {Nugent}, P.~E., {Ellis}, R.~S., {Conley},
  A.~J., {Le Borgne}, D., {Carlberg}, R.~G., {Guy}, J., {Balam}, D., {Basa},
  S., {Fouchez}, D., {Hook}, I.~M., {Hsiao}, E.~Y., {Neill}, J.~D., {Pain}, R.,
  {Perrett}, K.~M., and {Pritchet}, C.~J., ``{The type Ia supernova SNLS-03D3bb
  from a super-Chandrasekhar-mass white dwarf star},'' {\em Nature}~{\bf 443},
  308--311 (Sept. 2006).

\bibitem{Ngeow13}
{Ngeow}, C.-C., {Konidaris}, N., {Quimby}, R., {Ritter}, A., {Rudy}, A.~R.,
  {Lin}, E., and {Ben-Ami}, S., ``{The SED Machine: A Spectrograph to
  Efficiently Classify Transient Events Discovered by PTF},'' in [{\em IAU
  Symposium}{\nolinebreak\hspace{0.1em}]},  {Zhang}, C.~M., {Belloni}, T.,
  {M{\'e}ndez}, M., and {Zhang}, S.~N., eds., {\em IAU Symposium} {\bf 290},
  281--282 (Feb. 2013).

\bibitem{Nugent11}
{Nugent}, P.~E., {Sullivan}, M., {Cenko}, S.~B., {Thomas}, R.~C., {Kasen}, D.,
  {Howell}, D.~A., {Bersier}, D., {Bloom}, J.~S., {Kulkarni}, S.~R.,
  {Kandrashoff}, M.~T., {Filippenko}, A.~V., {Silverman}, J.~M., {Marcy},
  G.~W., and {Howard}, A.~W., ``{Supernova SN 2011fe from an exploding
  carbon-oxygen white dwarf star},'' {\em Nature}~{\bf 480},  344--347 (Dec.
  2011).

\bibitem{Woosley93}
{Woosley}, S.~E., {Langer}, N., and {Weaver}, T.~A., ``{The evolution of
  massive stars including mass loss - Presupernova models and explosion},''
  {\em \apj}~{\bf 411},  823--839 (July 1993).

\bibitem{MacFadyen99}
{MacFadyen}, A.~I. and {Woosley}, S.~E., ``{Collapsars: Gamma-Ray Bursts and
  Explosions in ``Failed Supernovae''},'' {\em \apj}~{\bf 524},  262--289 (Oct.
  1999).

\bibitem{Iwamoto98}
{Iwamoto}, K., {Mazzali}, P.~A., {Nomoto}, K., {Umeda}, H., {Nakamura}, T.,
  {Patat}, F., {Danziger}, I.~J., {Young}, T.~R., {Suzuki}, T., {Shigeyama},
  T., {Augusteijn}, T., {Doublier}, V., {Gonzalez}, J.-F., {Boehnhardt}, H.,
  and {Brewer}, J., ``{A hypernova model for the supernova associated with the
  {$\gamma$}-ray burst of 25 April 1998},'' {\em \nat}~{\bf 395},  672--674
  (Oct. 1998).

\bibitem{Mazzali03}
{Mazzali}, P.~A., {Deng}, J., {Tominaga}, N., {Maeda}, K., {Nomoto}, K.,
  {Matheson}, T., {Kawabata}, K.~S., {Stanek}, K.~Z., and {Garnavich}, P.~M.,
  ``{The Type Ic Hypernova SN 2003dh/GRB 030329},'' {\em \apjl}~{\bf 599},
  L95--L98 (Dec. 2003).

\bibitem{Mazzali05b}
{Mazzali}, P.~A., {Kawabata}, K.~S., {Maeda}, K., {Nomoto}, K., {Filippenko},
  A.~V., {Ramirez-Ruiz}, E., {Benetti}, S., {Pian}, E., {Deng}, J., {Tominaga},
  N., {Ohyama}, Y., {Iye}, M., {Foley}, R.~J., {Matheson}, T., {Wang}, L., and
  {Gal-Yam}, A., ``{An Asymmetric Energetic Type Ic Supernova Viewed Off-Axis,
  and a Link to Gamma Ray Bursts},'' {\em Science}~{\bf 308},  1284--1287 (May
  2005).

\bibitem{Eichler99}
{Eichler}, D. and {Levinson}, A., ``{Shading and Smothering of Gamma-Ray
  Bursts},'' {\em \apjl}~{\bf 521},  L117--L120 (Aug. 1999).

\bibitem{Gendre13}
{Gendre}, B., {Stratta}, G., {Atteia}, J.~L., {Basa}, S., {Bo{\"e}r}, M.,
  {Coward}, D.~M., {Cutini}, S., {D'Elia}, V., {Howell}, E.~J., {Klotz}, A.,
  and {Piro}, L., ``{The Ultra-long Gamma-Ray Burst 111209A: The Collapse of a
  Blue Supergiant?},'' {\em \apj}~{\bf 766},  30 (Mar. 2013).

\bibitem{Guidorzi06}
{Guidorzi}, C., {Monfardini}, A., {Gomboc}, A., {Mottram}, C.~J., {Mundell},
  C.~G., {Steele}, I.~A., {Carter}, D., {Bode}, M.~F., {Smith}, R.~J.,
  {Fraser}, S.~N., {Burgdorf}, M.~J., and {Newsam}, A.~M., ``{The Automatic
  Real-Time Gamma-Ray Burst Pipeline of the 2 m Liverpool Telescope},'' {\em
  \pasp}~{\bf 118},  288--296 (Feb. 2006).

\bibitem{Gotz09}
{G{\"o}tz}, D., {Paul}, J., {Basa}, S., {Wei}, J., {Zhang}, S.~N., {Atteia},
  J.-L., {Barret}, D., {Cordier}, B., {Claret}, A., {Deng}, J., {Fan}, X.,
  {Hu}, J.~Y., {Huang}, M., {Mandrou}, P., {Mereghetti}, S., {Qiu}, Y., and
  {Wu}, B., ``{SVOM: a new mission for Gamma-Ray Burst Studies},'' in [{\em
  American Institute of Physics Conference
  Series}{\nolinebreak\hspace{0.1em}]},  {Meegan}, C., {Kouveliotou}, C., and
  {Gehrels}, N., eds., {\em American Institute of Physics Conference Series}
  {\bf 1133},  25--30 (May 2009).

\bibitem{Aasi13}
{The LIGO Scientific Collaboration}, {the Virgo Collaboration}, {Aasi}, J.,
  {Abadie}, J., {Abbott}, B.~P., {Abbott}, R., {Abbott}, T., {Abernathy},
  M.~R., {Accadia}, T., {Acernese}, F., and et~al., ``{First Searches for
  Optical Counterparts to Gravitational-wave Candidate Events},'' {\em ArXiv
  e-prints}  (Oct. 2013).

\bibitem{Abadie10}
{Abadie}, J., {Abbott}, B.~P., {Abbott}, R., {Abernathy}, M., {Accadia}, T.,
  {Acernese}, F., {Adams}, C., {Adhikari}, R., {Ajith}, P., {Allen}, B., and
  et~al., ``{TOPICAL REVIEW: Predictions for the rates of compact binary
  coalescences observable by ground-based gravitational-wave detectors},'' {\em
  Classical and Quantum Gravity}~{\bf 27},  173001 (Sept. 2010).

\bibitem{VanEerten11}
{van Eerten}, H.~J. and {MacFadyen}, A.~I., ``{Synthetic Off-axis Light Curves
  for Low-energy Gamma-Ray Bursts},'' {\em \apjl}~{\bf 733},  L37 (June 2011).

\bibitem{Metzger12}
{Metzger}, B.~D. and {Berger}, E., ``{What is the Most Promising
  Electromagnetic Counterpart of a Neutron Star Binary Merger?},'' {\em
  \apj}~{\bf 746},  48 (Feb. 2012).

\bibitem{Berger13}
{Berger}, E., {Fong}, W., and {Chornock}, R., ``{An r-process Kilonova
  Associated with the Short-hard GRB 130603B},'' {\em \apjl}~{\bf 774},  L23
  (Sept. 2013).

\bibitem{Tanvir13}
{Tanvir}, N.~R., {Levan}, A.~J., {Fruchter}, A.~S., {Hjorth}, J., {Hounsell},
  R.~A., {Wiersema}, K., and {Tunnicliffe}, R.~L., ``{A `kilonova' associated
  with the short-duration {$\gamma$}-ray burst GRB130603B},'' {\em \nat}~{\bf
  500},  547--549 (Aug. 2013).

\bibitem{Aasi13b}
{LIGO Scientific Collaboration}, {Virgo Collaboration}, {Aasi}, J., {Abadie},
  J., {Abbott}, B.~P., {Abbott}, R., {Abbott}, T.~D., {Abernathy}, M.,
  {Accadia}, T., {Acernese}, F., and et~al., ``{Prospects for Localization of
  Gravitational Wave Transients by the Advanced LIGO and Advanced Virgo
  Observatories},'' {\em ArXiv e-prints}  (Apr. 2013).

\bibitem{Abadie12b}
{Abadie}, J., {Abbott}, B.~P., {Abbott}, R., {Abbott}, T.~D., {Abernathy}, M.,
  {Accadia}, T., {Acernese}, F., {Adams}, C., {Adhikari}, R., {Affeldt}, C.,
  and et~al., ``{First low-latency LIGO+Virgo search for binary inspirals and
  their electromagnetic counterparts},'' {\em \aap}~{\bf 541},  A155 (May
  2012).

\bibitem{Lorimer07}
{Lorimer}, D.~R., {Bailes}, M., {McLaughlin}, M.~A., {Narkevic}, D.~J., and
  {Crawford}, F., ``{A Bright Millisecond Radio Burst of Extragalactic
  Origin},'' {\em Science}~{\bf 318},  777-- (Nov. 2007).

\bibitem{Thornton13}
{Thornton}, D., {Stappers}, B., {Bailes}, M., {Barsdell}, B., {Bates}, S.,
  {Bhat}, N.~D.~R., {Burgay}, M., {Burke-Spolaor}, S., {Champion}, D.~J.,
  {Coster}, P., {D'Amico}, N., {Jameson}, A., {Johnston}, S., {Keith}, M.,
  {Kramer}, M., and {Levin}, L., ``{A Population of Fast Radio Bursts at
  Cosmological Distances},'' {\em Science}~{\bf 341},  53--56 (July 2013).

\end{thebibliography}
\bibliographystyle{spiebib2}   

\end{document}